# Ψ − MOYAL EQUATION

**E.E. Perepelkin[a,b,d], B.I. Sadovnikov[a], N.G. Inozemtseva[b,c], A.A. Korepanova[a]**

[a] *Faculty of Physics, Lomonosov Moscow State University, Moscow, 119991 Russia*
[b] *Moscow Technical University of Communications and Informatics, Moscow, 123423 Russia*
[c] *Dubna State University, Moscow region, Dubna,141980 Russia*
[d] *Joint Institute for Nuclear Research, Moscow region, Dubna,141980 Russia*

**Abstract**

A full consideration of classical and quantum systems with radiation (electromagnetic/gravitational) requires the involvement of a mathematical description in the generalized phase space of high kinematical values. Based on the dispersion chain of equations of quantum mechanics, we construct a generalization of the von Neumann equation for the density matrix in the phase space of fourth-order kinematical values. The paper introduces a new extended definition of the fourth rank Wigner function, which is constructed from the wave functions of the second rank. A new extended Moyal equation (PSI-Moyal equation) for the Wigner function of the fourth rank is obtained. Theorems on the properties of the new PSI-Moyal equation and its solutions are proved. An example of a model quantum system is considered in detail.

**Introduction**

The description of quantum systems in phase space using the Wigner function [1, 2] is widely used in quantum optics [3-5], quantum tomography [6,7], and quantum cryptography [8, 9]. In comparison with classical physical systems, quantum systems are characterized by a special «sensitivity», requiring for their full description to employ the mathematical apparatus of quantum mechanics of high kinematical values.

As is known, classical mechanics is built on second-order differential equations of motion. Quantum mechanics was built in a phenomenological way on the basis of analogues of the Hamiltonians of classical mechanics. With the development of physics in 1897, H. Lorentz [10] proposed a new equation of motion describing the radiation of an electromagnetic wave by a charged particle moving with acceleration. The Lorentz equation was a third-order equation, that is, it contained high-order kinematical value $\dddot{v}$ responsible for the power of electromagnetic radiation. Of course, Newton's mechanics could not contain information about electromagnetic radiation, since such processes were not experimentally and theoretically studied during its construction. Consideration of physical systems at the micro level has led to a significant rethinking of physical reality and the construction of quantum physics, *but* based on classical mechanics.

It seems natural that when describing physical systems at the micro level, it is necessary to take into account high kinematical values. In 1850, M. Ostrogradsky proposed the Lagrangian and Hamiltonian formalism of the mechanics of high kinematical values [11-13], which was subsequently applied in the field theory [14-17]. In the twentieth century, A. Vlasov introduced distribution functions in an infinite-dimensional generalized phase space [18, 19] and obtained a chain of equations for them. Despite the Heisenberg uncertainty principle, the first attempts to expand the consideration of quantum systems in the phase space were made in the works of E.

Wigner and Weil [1,2]. The Wigner function introduced in a phenomenological way became a quasi-probability density function due to its negative values [20-23]. The negative values of the probabilities of the Wigner function in a sense demonstrate that it has «non-classical», quantum nature. The phase space contains the coordinate and momentum, and the radiation processes are characterized by kinematic values of high orders ($\ddot{v}$).

In [24, 25, 26], the description of classical and quantum systems was considered in the framework of the dispersion chain of equations of quantum mechanics. The dispersion chain of equations of quantum mechanics [26] works with classical, that is, with positive probability density functions. The existence of a quasi-probability density function (Wigner function) suggests the idea of constructing a generalized analogue of the Wigner function for high kinematical values.

The aim of this paper is to introduce a new extension of the Wigner function for kinematical values $\vec{r}, \vec{v}, \dot{\vec{v}}, \ddot{\vec{v}}, \dddot{\vec{v}}$ and to obtain for it an extended analogue of the Moyal equation ($\Psi$-Moyal equation).

The paper has the following structure. §1 considers the construction of an extended analogue of the von Neumann equation for the density matrix of a quantum system described by kinematical values of up to and including the fourth order. The construction of the equation for the density matrix is based on the Schrödinger equation of the second rank [26] for wave function $\Psi(\vec{r}, \vec{v}, t)$. In §2 the definition of Wigner function $W^{1,2,3,4}(\vec{r}, \vec{v}, \dot{\vec{v}}, \ddot{\vec{v}}, t)$ of the forth rank is introduced, its properties are investigated as well as the ones of third-rank functions $W^{1,2,3}(\vec{r}, \vec{v}, \dot{\vec{v}}, t)$ and $W^{1,2,4}(\vec{r}, \vec{v}, \ddot{\vec{v}}, t)$. On the basis of the extended analogue of the von Neumann equation (§1) in §2, the $\Psi$-Moyal equation is obtained, which is satisfied by the function $W^{1,2,3,4}$. Equations are constructed for functions $W^{1,2,3}$ and $W^{1,2,4}$. The equation for function $W^{1,2}(\vec{r}, \vec{v}, t)$ coincides with the well-known Moyal equation. An analogue of the Vlasov-Moyal approximation [28] for mean kinematical value $\langle \dddot{\vec{v}} \rangle_{1,2,3,4}$ is suggested. In §3, a model example of constructing Wigner functions $W^{1,2,3,4}$, $W^{1,2,3}$ and $W^{1,2,4}$ for a quantum harmonic oscillator is given and the corresponding $\Psi$-Moyal equations are considered. The conclusions contain the main findings of the paper. Appendices A, B provide proofs of theorems and intermediate mathematical transformations.

## §1 The Von Neumann equation of the second rank

According to [26], the second-rank Schrödinger equation from the first group of the dispersion chain of equations of quantum mechanics has the form:

$$\frac{i}{\beta_n} \partial_n \Psi^{n,n+1} = \hat{H}_{n,n+1} \Psi^{n,n+1}, \quad (1.1)$$

$$\hat{H}_{n,n+1} = \hat{T}_{n,n+1} + U^{\widehat{n,n+1}} = -\alpha_{n+1}\beta_{n+1} \hat{p}_{n+1}^2 + U^{\widehat{n,n+1}}, \quad (1.2)$$

$$\partial_n = \frac{\partial}{\partial t} + \vec{\xi}^{n+1} \nabla_{\xi^n}, \quad \hat{p}_{n+1} = -\frac{i}{\beta_{n+1}} \nabla_{\xi^{n+1}}, \quad \alpha_{n+1} = -\frac{\hbar_{n+1}}{2m}, \quad \beta_{n+1} = \frac{1}{\hbar_{n+1}}, \quad (1.4)$$

where $\vec{\xi}^n$ is kinematical value of order $n$, belonging to generalized phase space $\Omega$ [27, 24]; $\partial_0 = \frac{\partial}{\partial t}$; $\hbar_n$, $m$ are constant. Operator $U^{\widehat{n,n+1}}$ corresponds to the generalized potential [26].

**Definition 1** *Let us put that equation (1.1) is written in the Vlasov form and operator (1.2) is denoted as* $\hat{H}^V_{n,n+1} \stackrel{det}{=} \hat{H}_{n,n+1}$. *Equation (1.1) can be rewritten in the form:*

$$\frac{i}{\beta_n}\partial_0 \Psi^{n,n+1} = \hat{H}^S_{n,n+1} \Psi^{n,n+1}, \qquad (1.5)$$

$$\hat{H}^S_{n,n+1} \stackrel{det}{=} -\alpha_{n+1}\beta_{n+1}\hat{p}^2_{n+1} + \vec{\xi}^{n+1}\hat{p}_n + U^{\widehat{n,n+1}}, \qquad (1.6)$$

*which we will call the Schrödinger form.*

Without loss of generality, we consider the case of $n = 1$. Operator $\hat{H}^S_{1,2}$ is not self-adjoint unlike operator $\hat{H}^V_{1,2}$. Let potential $U^{1,2}$ be stationary and operator (1.6) have its own orthonormal system of basis functions $\psi_n(\vec{r},\vec{v})$ with eigenvalues $E^{1,2}_n$:

$$\begin{aligned}\hat{H}^S_{1,2}\psi_n &= E^{1,2}_n \psi_n, \\ \overline{\hat{H}}^S_{1,2}\overline{\psi}_n &= \overline{E}^{1,2}_n \overline{\psi}_n.\end{aligned} \qquad (1.7)$$

Let us consider wave function $\Psi^{1,2}$ of the second rank satisfying equation (1.5):

$$i\hbar_2 \partial_0 \Psi^{1,2} = \hat{H}^S_{1,2} \Psi^{1,2}, \qquad (1.8)$$

and allowing an expansion of the form:

$$\Psi^{1,2}(\vec{r},\vec{v},t) = \sum_n C_n(t)\psi_n(\vec{r},\vec{v}), \qquad (1.9)$$

$$C_n(t) = c_n e^{-i\frac{E^{1,2}_n}{\hbar_2}t},$$

where $c_n$ are constant values. Substituting function (1.9) into equation (1.8) leads to equation (1.7).

**Lemma** *Let wave function (1.9) be put into correspondence with a matrix*

$$\rho_{kn} \stackrel{det}{=} \overline{C}_n(t)C_k(t), \qquad (1.10)$$

*then an equation is true:*

$$\partial_0 \hat{\rho} = \frac{i}{\hbar_2}\left[\hat{\rho}, \hat{H}^S_{1,2}\right]^* \stackrel{det}{=} \frac{i}{\hbar_2}\left(\hat{\rho}\overline{\hat{H}}^S_{1,2} - \hat{H}^S_{1,2}\hat{\rho}\right). \qquad (1.11)$$

The proof of Lemma is given in Appendix A.

Equation (1.11) can be interpreted as an analogue of the von Neumann equation for the density matrix of a quantum system.

## §2 Generalized Wigner function

If we introduce the Wigner function of high kinematical values, then equation (1.11) will make it possible to generalize the Moyal equation.

**Definition 2** *For wave function $\Psi^{1,2}(\vec{r},\vec{v},t)$ of the second rank (1.9) we define Wigner functions of the fourth rank $W^{1,2,3,4} = W^{1,2,3,4}(\vec{r},\vec{p},\dot{\vec{p}},\ddot{\vec{p}},t)$ and the third rank $W^{1,2,3} = W^{1,2,3}(\vec{r},\vec{p},\dot{\vec{p}},t)$, $W^{1,2,4} = W^{1,2,4}(\vec{r},\vec{p},\ddot{\vec{p}},t)$ as:*

$$W^{1,2,3,4} \stackrel{det}{=} \frac{1}{(2\pi\hbar_2)^6} \int_{(\infty)}\int_{(\infty)} \bar{\Psi}^{1,2}(\vec{r}',\vec{v}',t)\Psi^{1,2}(\vec{r}'',\vec{v}'',t)\exp\left(i\frac{\vec{s}_1\ddot{\vec{p}} - \vec{s}_2\dot{\vec{p}}}{\hbar_2}\right)d^3s_1 d^3s_2, \qquad (2.1)$$

$$W^{1,2,3} \stackrel{det}{=} \frac{1}{(2\pi\hbar_2)^3} \int_{(\infty)} \bar{\Psi}^{1,2}(\vec{r},\vec{v}',t)\Psi^{1,2}(\vec{r},\vec{v}'',t)\exp\left(-i\frac{\vec{s}_2\dot{\vec{p}}}{\hbar_2}\right)d^3s_2, \qquad (2.2)$$

$$W^{1,2,4} \stackrel{det}{=} \frac{1}{(2\pi\hbar_2)^3} \int_{(\infty)} \bar{\Psi}^{1,2}(\vec{r}',\vec{v},t)\Psi^{1,2}(\vec{r}'',\vec{v},t)\exp\left(i\frac{\vec{s}_1\ddot{\vec{p}}}{\hbar_2}\right)d^3s_1, \qquad (2.3)$$

*where the following notations are used:* $\vec{p} = m\vec{v}$, $\dot{\vec{p}} = m\dot{\vec{v}}$, $\ddot{\vec{p}} = m\ddot{\vec{v}}$,

$$\vec{r}' = \vec{r} - \frac{\vec{s}_1}{2},\ \vec{v}' = \vec{v} - \frac{\vec{s}_2}{2},\ \vec{r}'' = \vec{r} + \frac{\vec{s}_1}{2},\ \vec{v}'' = \vec{v} + \frac{\vec{s}_2}{2}.$$

Functions (1.12) and (1.13) possess the properties (see Appendix A):

$$W^{1,2,3} = \int_{(\infty)} W^{1,2,3,4} d^3\ddot{p}, \qquad |\Psi^{1,2}|^2 = \int_{(\infty)} W^{1,2,3} d^3\dot{p}, \qquad (2.4)$$

$$W^{1,2,4} = \int_{(\infty)} W^{1,2,3,4} d^3\dot{p}, \qquad |\Psi^{1,2}|^2 = \int_{(\infty)} W^{1,2,4} d^3\ddot{p}.$$

**Theorem 1** *Let the conditions of the lemma for the wave function $\Psi^{1,2}(\vec{r},\vec{v},t)$ (1.9) be satisfied, and potential admit $U^{1,2}$ expansion in a Taylor series in respect to variables $\vec{r}$ and $\vec{v}$, then Wigner function of the fourth rank (1.12) satisfies the equation:*

$$\partial_{1,2,3}W^{1,2,3,4} = \left(\frac{\partial}{\partial t} + \frac{\vec{p}}{m}\nabla_r + \dot{\vec{p}}\nabla_p + \ddot{\vec{p}}\nabla_{\dot{p}}\right)W^{1,2,3,4} =$$
$$= \sum_{n=0}^{+\infty}\sum_{l=0}^{+\infty} \frac{(-1)^{n+l}(\hbar_2/2)^{2l}m^{2l-n+1}}{n!(2l-n+1)!} U^{1,2}\left(\bar{\nabla}_r,\bar{\nabla}_{\ddot{p}}\right)^n \left(\bar{\nabla}_p,\bar{\nabla}_{\dot{p}}\right)^{2l-n+1} W^{1,2,3,4}. \qquad (2.5)$$

The proof of Theorem 1 is given in Appendix A.

Equation (2.5) is an analogue of the Moyal equation for Wigner function $W^{1,2} \stackrel{det}{=} W(\vec{r}, \vec{p}, t)$ in the phase space. Let us consider some particular cases of equation (2.5). At $l = 0$ (indices $n = 0, 1$), on the right side of equation (2.5) there will be no summands with multipliers $\hbar_2$, so we move them to the left:

$$\left[\frac{\partial}{\partial t} + \frac{\vec{p}}{m}\nabla_r + \dot{\vec{p}}\nabla_p + \left(\ddot{\vec{p}} - \nabla_v U^{1,2}\right)\nabla_{\dot{p}} + \nabla_r U^{1,2}\nabla_{\ddot{p}}\right] W^{1,2,3,4} =$$
$$= \sum_{n=0}^{+\infty}\sum_{l=1}^{+\infty} \frac{(-1)^{n+l}(\hbar_2/2)^{2l} m^{2l-n+1}}{n!(2l-n+1)!} U^{1,2} \left(\vec{\nabla}_r, \vec{\nabla}_{\ddot{p}}\right)^n \left(\vec{\nabla}_p, \vec{\nabla}_{\dot{p}}\right)^{2l-n+1} W^{1,2,3,4}, \quad (2.6)$$

or

$$\left[\frac{\partial}{\partial t} + \vec{v}\nabla_r + \dot{\vec{v}}\nabla_v + \left(\ddot{\vec{v}} - \frac{1}{m}\nabla_v U^{1,2}\right)\nabla_{\dot{v}} + \frac{1}{m}\nabla_r U^{1,2}\nabla_{\ddot{v}}\right] W^{1,2,3,4} =$$
$$= \frac{1}{m}\sum_{n=0}^{+\infty}\sum_{l=1}^{+\infty} \frac{(-1)^{n+l}(\hbar_2/2m)^{2l}}{n!(2l-n+1)!} U^{1,2} \left(\vec{\nabla}_r, \vec{\nabla}_{\ddot{v}}\right)^n \left(\vec{\nabla}_v, \vec{\nabla}_{\dot{v}}\right)^{2l-n+1} W^{1,2,3,4}, \quad (2.7)$$

where $W^{1,2,3,4} = W^{1,2,3,4}(\vec{r}, \vec{v}, \dot{\vec{v}}, \ddot{\vec{v}}, t) = W^{1,2,3,4}(\vec{r}, \vec{p}, \dot{\vec{p}}, \ddot{\vec{p}}, t)$. If potential $U^{1,2}(\vec{r}, \vec{v}) = U_1(\vec{r})$ is independent of the velocity, then equation (2.7) will take the form:

$$\left[\frac{\partial}{\partial t} + \vec{v}\nabla_r + \dot{\vec{v}}\nabla_v + \ddot{\vec{v}}\nabla_{\dot{v}} + \frac{1}{m}\nabla_r U_1 \nabla_{\ddot{v}}\right] W^{1,2,3,4} = \frac{1}{m}\sum_{l=1}^{+\infty} \frac{(-1)^{l+1}(\hbar_2/2m)^{2l}}{(2l+1)!} U_1 \left(\vec{\nabla}_r, \vec{\nabla}_{\ddot{v}}\right)^{2l+1} W^{1,2,3,4}, \quad (2.8)$$

where it is taken into account that $n = 2l + 1$. Equation (2.8) can be written as the fourth Vlasov equation, if a new approximation of mean kinematical $\langle \dddot{\vec{v}} \rangle$ is introduced for the chain break.

**Definition 3** *Let us define the second Vlasov-Moyal approximation for mean kinematical $\langle \dddot{\vec{v}} \rangle_{1,2,3,4}$ with potential $U^{1,2}$ as:*

$$\langle \dddot{v}_\mu \rangle_{1,2,3,4} = \frac{1}{m}\sum_{l=0}^{+\infty} \frac{(-1)^l(\hbar_2/2m)^{2l}}{(2l+1)!} \frac{\partial^{2l+1} U^{1,2}}{\partial x_\mu^{2l+1}} \frac{1}{f^{1,2,3,4}} \frac{\partial^{2l} f^{1,2,3,4}}{\partial \ddot{v}_\mu^{2l}}, \quad (2.9)$$

$$\langle \dddot{v}_\mu \rangle_{1,2,4} = \frac{1}{m}\sum_{l=0}^{+\infty} \frac{(-1)^l(\hbar_2/2m)^{2l}}{(2l+1)!} \frac{\partial^{2l+1} U^{1,2}}{\partial x_\mu^{2l+1}} \frac{1}{f^{1,2,4}} \frac{\partial^{2l} f^{1,2,4}}{\partial \ddot{v}_\mu^{2l}},$$

where $\mu = 1, 2, 3$; $f^{1,2,3,4}\langle \dddot{\vec{v}} \rangle_{1,2,3,4} \stackrel{det}{=} \int_{(\infty)} \dddot{\vec{v}} f^{1,2,3,4,5} d^3\dddot{v}$, $f^{1,2,4}\langle \dddot{\vec{v}} \rangle_{1,2,4} \stackrel{det}{=} \int_{(\infty)} \langle \dddot{\vec{v}} \rangle_{1,2,3,4} f^{1,2,3,4} d^3\dot{v}$.

**Theorem 2** *The fourth Vlasov equation for distribution function $f^{1,2,3,4}$*

$$\frac{\partial f^{1,2,3,4}}{\partial t} + \mathrm{div}_r\left[f^{1,2,3,4}\vec{v}\right] + \mathrm{div}_v\left[f^{1,2,3,4}\dot{\vec{v}}\right] + \mathrm{div}_{\dot{v}}\left[f^{1,2,3,4}\ddot{\vec{v}}\right] + \mathrm{div}_{\ddot{v}}\left[f^{1,2,3,4}\langle \dddot{\vec{v}} \rangle_{1,2,3,4}\right] = 0, \quad (2.10)$$

*with the second Vlasov-Moyal approximation (2.9) goes into the second Moyal equation (2.8) with potential $U^{1,2}(\vec{r},\vec{v}) = U_1(\vec{r})$ for function $W^{1,2,3,4}$ and in this case*

$$\langle \ddot{v}_\mu \rangle_{1,2,3} = \langle \ddot{v}_\mu \rangle_{1,2} = \langle \ddot{v}_\mu \rangle_1 = \frac{1}{m} \frac{\partial U_1}{\partial x_\mu}. \quad (2.11)$$

The proof of Theorem 2 is given in Appendix A.

**Theorem 3** *Let the equation (2.7) be satisfied with potential $U^{1,2}(\vec{r},\vec{v})$ depending on the coordinate and velocity, then Wigner functions of the third rank $W^{1,2,3}$, $W^{1,2,4}$ and the second rank $W^{1,2}$ satisfy the equations:*

$$\frac{\partial W^{1,2,3}}{\partial t} + \text{div}_r \left[ W^{1,2,3} \vec{v} \right] + \text{div}_v \left[ W^{1,2,3} \dot{\vec{v}} \right] + \text{div}_{\dot{v}} \left[ W^{1,2,3} \langle \ddot{\vec{v}} \rangle_{1,2,3} \right] = \sum_{l=0}^{+\infty} \frac{(-1)^l (\hbar_2/2m)^{2l}}{m(2l+1)!} U^{1,2} \left( \bar{\nabla}_v, \bar{\nabla}_{\dot{v}} \right)^{2l+1} W^{1,2,3},$$

(2.12)

$$\frac{\partial W^{1,2,4}}{\partial t} + \text{div}_r \left[ W^{1,2,4} \vec{v} \right] + \text{div}_v \left[ W^{1,2,4} \langle \dot{\vec{v}} \rangle_{1,2,4} \right] + \text{div}_{\dot{v}} \left[ W^{1,2,4} \langle \ddot{\vec{v}} \rangle_{1,2,4} \right] = 0, \quad (2.13)$$

$$\frac{\partial W^{1,2}}{\partial t} + \text{div}_r \left[ W^{1,2} \vec{v} \right] + \text{div}_v \left[ W^{1,2} \langle \dot{\vec{v}} \rangle_{1,2} \right] = 0, \quad (2.14)$$

*where*

$$W^{1,2} \langle \dot{\vec{v}} \rangle_{1,2} = m \int_{(\infty)} \dot{\vec{v}} \, W^{1,2,3} \, d^3\dot{v}, \quad W^{1,2,4} \langle \dot{\vec{v}} \rangle_{1,2,4} = m \int_{(\infty)} \dot{\vec{v}} \, W^{1,2,3,4} \, d^3\dot{v}, \quad (2.15)$$

$$W^{1,2,3} \langle \ddot{\vec{v}} \rangle_{1,2,3} = m \int_{(\infty)} \ddot{\vec{v}} \, W^{1,2,3,4} \, d^3\ddot{v},$$

*and mean flow $\langle \ddot{\vec{v}} \rangle_{1,2,4}$ satisfies the second Vlasov-Moyal approximation (2.9).*

The proof of Theorem 3 is given in Appendix A.

Equations (2.13) and (2.14) coincide with the equations from the Vlasov dispersion chain for distribution functions $f^{1,2,4}$ and $f^{1,2}$, respectively. Equation (2.14) is actually the Moyal equation in the first Vlasov-Moyal approximation [28]

$$\langle \dot{v}_\mu \rangle_{1,2} = \sum_{l=0}^{+\infty} \frac{(-1)^{l+1} (\hbar_2/2m)^{2l}}{m(2l+1)!} \frac{\partial^{2l+1} U^1}{\partial x_\mu^{2l+1}} \frac{1}{f^{1,2}} \frac{\partial^{2l} f^{1,2}}{\partial v_\mu^{2l}}, \quad (2.16)$$

where $U^1$ is the potential from the Schrödinger equation. Equation (2.14) is natural for function $W^{1,2}$, since, by property (2.4) $W^{1,2} = |\Psi^{1,2}|^2$, it automatically satisfies the second Vlasov equation. Equation (2.13), according to the dispersion chain of Vlasov equations, has two sources of dissipations [25] due to the values

$$Q_{1,2,4}^2 = \text{div}_v \langle \dot{\vec{v}} \rangle_{1,2,4}, \quad Q_{1,2,4}^4 = \text{div}_{\dot{v}} \langle \ddot{\vec{v}} \rangle_{1,2,4}, \quad Q_{1,2}^2 = \text{div}_v \langle \dot{\vec{v}} \rangle_{1,2}. \quad (2.17)$$

It can be seen from expressions (2.9) and (2.16) that, in the general case, the sources of dissipations (2.17) are nonzero, therefore, probability densities $W^{1,2,4}$ and $W^{1,2}$ will not be constant along phase trajectories (2.13), (2.14):

$$\hat{\pi}_{1,2,4} S^{1,2,4} = -\left(Q^2_{1,2,4} + Q^4_{1,2,4}\right), \quad (2.18)$$
$$\hat{\pi}_{1,2} S^{1,2} = -Q^2_{1,2},$$

where $S^{1,2,4} = \operatorname{Ln} W^{1,2,4}$, $S^{1,2} = \operatorname{Ln} W^{1,2}$; and operator $\hat{\pi}$ is of the form [25]:

$$\hat{\pi}_{1,2} = \partial_0 + \vec{v}\nabla_r + \left\langle\dot{\vec{v}}\right\rangle_{1,2} \nabla_v,$$
$$\hat{\pi}_{1,2,4} = \partial_0 + \vec{v}\nabla_r + \left\langle\dot{\vec{v}}\right\rangle_{1,2,4} \nabla_v + \left\langle\ddot{\vec{v}}\right\rangle_{1,2,4} \nabla_{\ddot{v}}. \quad (2.19)$$

Equation (2.12) differs significantly from the equation from the Vlasov dispersion chain by the presence of a nonzero right-hand side. As noted in Theorem 2, only for potential $U^{1,2}(\vec{r},\vec{v})$ that does not depend on the velocity does equation (2.12) go into the fourth Vlasov equation (2.10). Note that the chain of Vlasov equations is based on the first principle – the law of conservation of probabilities. The presence of the right-hand side in equation (2.12) indicates a violation of this principle, that is, the «openness» of the system. In the process of averaging equation (2.12) over spaces $\dot{\vec{v}}$ or $\ddot{\vec{v}}$, as can be seen from Theorem 3, information about the «openness» of the system is lost and the corresponding equations (2.13) and (2.14) illustrate the conservation of probability in phase subspaces $(\vec{r},\vec{v})$, $(\vec{r},\ddot{\vec{v}})$.

## §3 Example

In [26], wave function $\Psi^{1,2}$ of a quantum harmonic oscillator in the phase space was considered:

$$\Psi^{1,2}(x,v,t) = \sqrt{\frac{m}{\pi\hbar}} \cdot \exp\left[-\frac{1}{\hbar\omega}\left(\frac{mv^2}{2} + \frac{m\omega^2 x^2}{2}\right) - i\left(\frac{m\omega^2}{\hbar_2}xv + \frac{E^{1,2}}{\hbar_2}t\right)\right], \quad (3.1)$$

which satisfies the Schrödinger equation of the second rank (1.1)/(1.5) with the potential

$$U^{1,2} = E^{1,2} - \frac{\hbar_2^2}{2\hbar\omega} + m\omega^2\left(1 + \frac{\hbar_2^2}{2\hbar^2\omega^4}\right)v^2 - \frac{1}{2}m\omega^4 x^2, \quad (3.2)$$

where $E^{1,2} = \dfrac{\hbar_2 \omega_2}{2}$ is a constant value. According to property (2.4), Wigner function $W^{1,2}$ corresponds to wave function (3.1)

$$W^{1,2}(x,v,t) = \left|\Psi^{1,2}\right|^2 = W(x,mv) = \frac{m}{\pi\hbar} \cdot \exp\left[-\frac{m}{\hbar\omega}\left(v^2 + \omega^2 x^2\right)\right], \quad (3.3)$$

which satisfies the equation (2.14). Note that initial potential $U^1(x) = \dfrac{m\omega^2 x^2}{2}$ of the harmonic oscillator does not contribute to Vlasov approximation (2.16), which leads to the absence of sources of dissipations $Q^2_{1,2}$ (2.17), (2.18).

Wigner function $W^{1,2,3,4}$ (2.1) in the one-dimensional case for wave function (3.1) is of the form:

$$W^{1,2,3,4} = \frac{1}{(2\pi\hbar_2)^2}\int_{-\infty}^{+\infty}\int_{-\infty}^{+\infty}\overline{\Psi}^{1,2}\left(x-\frac{s_1}{2}, v-\frac{s_2}{2}\right)\Psi^{1,2}\left(x+\frac{s_1}{2}, v+\frac{s_2}{2}\right)\exp\left(i\frac{s_1\ddot{p}-s_2\dot{p}}{\hbar_2}\right)ds_1 ds_2. \quad (3.4)$$

The calculation of integral (3.4) leads to the expression (see Appendix B):

$$W^{1,2,3,4} = \frac{1}{(\pi\hbar_2)^2}\exp\left\{-\frac{m}{\hbar\omega}\left[\left(v^2+\omega^2 x^2\right)+\frac{1}{\omega^4}\left(\omega^2 v-\ddot{v}\right)^2+\frac{1}{\omega^2}\left(\omega^2 x+\dot{v}\right)^2\right]\right\}. \quad (3.5)$$

Wigner function (3.5) must satisfy the second Moyal equation (2.7) from Theorem 1:

$$\left[\frac{\partial}{\partial t}+v\frac{\partial}{\partial x}+\dot{v}\frac{\partial}{\partial v}+\left(\ddot{v}-3\omega^2 v\right)\frac{\partial}{\partial \dot{v}}-\omega^4 x\frac{\partial}{\partial \ddot{v}}\right]W^{1,2,3,4} = 0, \quad (3.4)$$

in which the form of potential $U^{1,2}$ (3.2) is taken into account. Direct substitution of expression (3.5) into equation (3.4) leads to the correct identity (see Appendix B).

Integration of function (3.5) over spaces $\ddot{p}$ and $\dot{p}$ gives Wigner functions of the third rank $W^{1,2,3}$ and $W^{1,2,4}$, respectively, (2.2)–(2.4) (see Appendix B):

$$W^{1,2,3} = \sqrt{\frac{m}{\pi^3\hbar^3\omega^3}}\exp\left\{-\frac{m}{\hbar\omega}\left[v^2+\omega^2 x^2+\left(\frac{\omega^2 x+\dot{v}}{\omega}\right)^2\right]\right\}, \quad (3.5)$$

$$W^{1,2,4} = \sqrt{\frac{m\omega}{\pi^3\hbar_2^3}}\exp\left\{-\frac{m}{\hbar\omega}\left[v^2+\omega^2 x^2+\left(\frac{\ddot{v}-\omega^2 v}{\omega^2}\right)^2\right]\right\}, \quad (3.6)$$

Repeated integration of functions (3.5) and (3.6) over spaces $\dot{p}$ and $\ddot{p}$, respectively, will lead to the same function $W^{1,2}$ (see Appendix B):

$$W^{1,2} = \frac{m}{\pi\hbar}\exp\left[-\frac{m}{\hbar\omega}\left(v^2+\omega^2 x^2\right)\right]. \quad (3.7)$$

Note that function (3.7) coincides with function (3.3). Function (3.5) must satisfy equation (2.12)

$$\left[\frac{\partial}{\partial t}+v\frac{\partial}{\partial x}+\dot{v}\frac{\partial}{\partial v}\right]W^{1,2,3}+\frac{\partial}{\partial \dot{v}}\left[W^{1,2,3}\langle\ddot{v}\rangle_{1,2,3}\right] = 3\omega^2 v\frac{\partial}{\partial \dot{v}}W^{1,2,3}, \quad (3.8)$$

where the form of potential $U^{1,2}$ on the right-hand side of the equation is taken into account. The mean value $\langle\ddot{v}\rangle_{1,2,3}$ in equation (3.8) can be found using function (3.5) by formula (2.15):

$$W^{1,2,3}\langle\ddot{v}\rangle_{1,2,3} = \sqrt{\frac{m\omega}{\pi^3\hbar^3}}\exp\left\{-\frac{m}{\hbar\omega}\left[v^2+\omega^2x^2+\left(\frac{\omega^2x+\dot{v}}{\omega}\right)^2\right]\right\}v = \omega^2v\,W^{1,2,3},$$

$$\langle\ddot{v}\rangle_{1,2,3} = \omega^2 v. \tag{3.9}$$

where, according to (3.5), random variable $\dot{v}$ has a normal distribution. We note an interesting property of expression (3.9). In [29], the second Vlasov approximation was obtained for the value $\langle\ddot{v}\rangle_{1,2,3}$:

$$\langle\ddot{v}\rangle_{1,2,3} = \frac{1}{m}\frac{\partial N_{1,2}}{\partial x}, \quad N_{1,2} = \left(\frac{\partial}{\partial t}+v\frac{\partial}{\partial x}\right)U^1(x,t), \tag{3.10}$$

where $N_{1,2}$ is the power of radiation. For potential $U^1(x) = \frac{m\omega^2x^2}{2}$ under consideration, expression (3.10) goes into representation (3.9).

Substituting expressions (3.9) and (3.5) into equation (3.8) gives the correct identity.

Similarly, for function (3.6), equation (2.13) takes the form:

$$\left[\frac{\partial}{\partial t}+v\frac{\partial}{\partial x}\right]W^{1,2,4}+\frac{\partial}{\partial v}\left[\langle\dot{v}\rangle_{1,2,4}W^{1,2,4}\right]+\frac{\partial}{\partial\dot{v}}\left[\langle\ddot{v}\rangle_{1,2,4}W^{1,2,4}\right]=0. \tag{3.11}$$

It follows from the second Vlasov-Moyal approximation (2.9) that the following representation is true for potential (3.2)

$$\langle\ddot{v}\rangle_{1,2,3,4} = \langle\ddot{v}\rangle_{1,2,4} = \frac{1}{m}\frac{\partial U^{1,2}}{\partial x} = -\omega^4 x. \tag{3.12}$$

The average acceleration flux $\langle\dot{v}\rangle_{1,2,4}$ can be found by the formula (2.15)

$$\langle\dot{v}\rangle_{1,2,4} = -\omega^2 x. \tag{3.13}$$

where it is taken into account that random $\dot{v}$ variable is distributed according to the normal law (3.5). Substituting expressions (3.12) and (3.13) into equation (3.11), we obtain

$$\left[\frac{\partial}{\partial t}+v\frac{\partial}{\partial x}-\omega^2x\frac{\partial}{\partial v}-\omega^4x\frac{\partial}{\partial\dot{v}}\right]W^{1,2,4}=0. \tag{3.14}$$

Direct substitution of function (3.6) into equation (3.14) leads to the correct identity.

The last equation (2.14) with approximation (2.16) is the Moyal equation for the Wigner function. Function $W^{1,2}$ (3.7) is the Wigner function of the ground state of the harmonic oscillator, therefore, $W^{1,2}$ satisfies the equation (2.14).

**Conclusions**

The main criterion for choosing the form of Wigner function (2.1) was the possibility of constructing $\Psi$-Moyal-equation (2.5) using von Neumann equation (1.11). As a result, the $\Psi$-

Moyal equation has a similar structure to the usual Moyal equation. The more complex structure of the right-hand side of the $\Psi$-Moyal equation in the general case does not make it possible to write it in the form of the fourth Vlasov equation (2.10). The exception is the case when $U^{1,2}(\vec{r},\vec{v}) = U_1(\vec{r})$. Note that the ordinary Moyal equation is a special case of the second Vlasov equation (2.14) with Vlasov-Moyal approximation (2.16). On the other hand, for Wigner function $W^{1,2,4}$, equation (2.13) is a particular case of the second-rank equation from the dispersion chain of Vlasov equations [25] with Vlasov-Moyal approximations (2.9) and (2.16) [28].

It turns out that to construct fourth rank Wigner function $W^{1,2,3,4}$, information from the usual phase space is sufficient, that is, knowledge of wave functions of the second rank $\Psi^{1,2}$. Functions of the third rank $W^{1,2,3}$, $W^{1,2,4}$ are obtained by integrating $W^{1,2,3,4}$ over the corresponding phase subspaces.

It seems interesting that the approximations of kinematical values $\langle \vec{v} \rangle_{1,2,3}$ (3.9) and (3.10), which characterize the radiation power [29], coincide. Note that approximation (3.10) was obtained without using the apparatus of the Wigner function from statistical physics.

**Acknowledgements**

This research has been supported by the Interdisciplinary Scientific and Educational School of Moscow University «Photonic and Quantum Technologies. Digital Medicine».

**Appendix A**

*Proof of Lemma*

The matrix of operator $\hat{H}^S_{1,2}$ in the basis of eigenfunctions $\psi_n$ has a diagonal form and the diagonal elements are eigenvalues $E_n^{1,2}$, i.e.

$$\mathrm{H}_{jn} = \int_{(\infty)}\int_{(\infty)} \overline{\psi}_j \hat{H}^S_{1,2} \psi_n d^3r d^3v = \int_{(\infty)}\int_{(\infty)} \overline{\psi}_j E_n^{1,2} \psi_n d^3r d^3v = E_n^{1,2} \delta_{jn}, \quad (A.1)$$

$$\overline{\mathrm{H}}_{jn} = \int_{(\infty)}\int_{(\infty)} \psi_j \overline{\hat{H}^S_{1,2}} \overline{\psi}_n d^3r d^3v = \int_{(\infty)}\int_{(\infty)} \psi_j \overline{E}_n^{1,2} \overline{\psi}_n d^3r d^3v = \overline{E}_n^{1,2} \delta_{jn}.$$

Let us differentiate matrix elements $\rho_{kn}$ with respect to time:

$$\partial_0 \rho_{kn} = \overline{c}_n c_k \frac{\partial}{\partial t} e^{i\frac{\overline{E}_n^{1,2} - E_k^{1,2}}{\hbar_2}t} = \overline{c}_n c_k e^{i\frac{\overline{E}_n^{1,2} - E_k^{1,2}}{\hbar_2}t} i\frac{\overline{E}_n^{1,2} - E_k^{1,2}}{\hbar_2} = \frac{i}{\hbar_2}\left(\overline{E}_n^{1,2} - E_k^{1,2}\right)\overline{C}_n C_k,$$

$$\partial_0 \rho_{kn} = \frac{i}{\hbar_2}\left(\overline{E}_n^{1,2} - E_k^{1,2}\right)\rho_{kn}. \quad (A.2)$$

Given expression (A.1), equation (A.2) takes the form:

$$\overline{E}_n^{1,2} \rho_{kn} = \sum_s \rho_{ks} \overline{E}_n^{1,2} \delta_{sn} = \sum_s \rho_{ks} \overline{\mathrm{H}}_{sn}, \quad (A.3)$$

$$E_k^{1,2} \rho_{kn} = \sum_s \rho_{sn} E_s^{1,2} \delta_{sk} = \sum_s \mathrm{H}_{ks} \rho_{sn},$$

as a result

$$\partial_0 \rho_{kn} = \frac{i}{\hbar_2} \sum_s \left( \rho_{ks} \bar{H}_{sn} - H_{ks} \rho_{sn} \right). \tag{A.4}$$

In the matrix/operator form, equation (A.3) can be written as

$$\partial_0 \hat{\rho} = \frac{i}{\hbar_2} \left[ \hat{\rho}, \hat{H}^S_{1,2} \right]^*, \tag{A.5}$$

where the asterisk sign indicates conjugacy in the commutator. The lemma is proven.

Let us calculate the integral of the Wigner function of the fourth and third ranks, we get:

$$\int_{(\infty)} W^{1\ldots 4} d^3\ddot{p} = \frac{1}{(2\pi\hbar_2)^6} \int_{(\infty)} e^{-i\frac{\vec{s}_2 \dot{\vec{p}}}{\hbar_2}} d^3s_2 \int_{(\infty)} \bar{\Psi}^{1,2}(\vec{r}',\vec{v}',t) \Psi^{1,2}(\vec{r}'',\vec{v}'',t) d^3s_1 \int_{(\infty)} e^{i\frac{\vec{s}_1 \ddot{\vec{p}}}{\hbar_2}} d^3\ddot{p} =$$

$$= \frac{1}{(2\pi\hbar_2)^3} \int_{(\infty)} e^{-i\frac{\vec{s}_2 \dot{\vec{p}}}{\hbar_2}} d^3s_2 \int_{(\infty)} \bar{\Psi}^{1,2}(\vec{r}',\vec{v}',t) \Psi^{1,2}(\vec{r}'',\vec{v}'',t) \delta(\vec{s}_1) d^3s_1 = \tag{A.6}$$

$$= \frac{1}{(2\pi\hbar_2)^3} \int_{(\infty)} e^{-i\frac{\vec{s}_2 \dot{\vec{p}}}{\hbar_2}} \bar{\Psi}^{1,2}(\vec{r},\vec{v}',t) \Psi^{1,2}(\vec{r},\vec{v}'',t) d^3s_2 = W^{1,2,3},$$

$$\int_{(\infty)} W^{1,2,3} d^3\dot{p} = \frac{1}{(2\pi\hbar_2)^3} \int_{(\infty)} \bar{\Psi}^{1,2}(\vec{r},\vec{v}',t) \Psi^{1,2}(\vec{r},\vec{v}'',t) d^3s_2 \int_{(\infty)} e^{-i\frac{\vec{s}_2 \dot{\vec{p}}}{\hbar_2}} d^3\dot{p} =$$

$$= \int_{(\infty)} \bar{\Psi}^{1,2}(\vec{r},\vec{v}',t) \Psi^{1,2}(\vec{r},\vec{v}'',t) \delta(\vec{s}_2) d^3s_2 = \bar{\Psi}^{1,2}(\vec{r},\vec{v},t) \Psi^{1,2}(\vec{r},\vec{v},t), \tag{A.7}$$

$$\int_{(\infty)} W^{1\ldots 4} d^3\dot{p} = \frac{1}{(2\pi\hbar_2)^6} \int_{(\infty)} d^3s_2 \int_{(\infty)} e^{i\frac{\vec{s}_1 \ddot{\vec{p}}}{\hbar_2}} \bar{\Psi}^{1,2}(\vec{r}',\vec{v}',t) \Psi^{1,2}(\vec{r}'',\vec{v}'',t) d^3s_1 \int_{(\infty)} e^{-i\frac{\vec{s}_2 \dot{\vec{p}}}{\hbar_2}} d^3\dot{p} =$$

$$= \frac{1}{(2\pi\hbar_2)^3} \int_{(\infty)} \bar{\Psi}^{1,2}(\vec{r}',\vec{v}',t) \Psi^{1,2}(\vec{r}'',\vec{v}'',t) \delta(\vec{s}_2) d^3s_2 \int_{(\infty)} e^{i\frac{\vec{s}_1 \ddot{\vec{p}}}{\hbar_2}} d^3s_1 =$$

$$= \frac{1}{(2\pi\hbar_2)^3} \int_{(\infty)} \bar{\Psi}^{1,2}(\vec{r}',\vec{v},t) \Psi^{1,2}(\vec{r}'',\vec{v},t) e^{i\frac{\vec{s}_1 \ddot{\vec{p}}}{\hbar_2}} d^3s_1 = W^{1,2,4},$$

$$\int_{(\infty)} W^{1,2,4} d^3\ddot{p} = \frac{1}{(2\pi\hbar_2)^3} \int_{(\infty)} \bar{\Psi}^{1,2}(\vec{r}',\vec{v},t) \Psi^{1,2}(\vec{r}'',\vec{v},t) d^3s_1 \int_{(\infty)} e^{i\frac{\vec{s}_1 \ddot{\vec{p}}}{\hbar_2}} d^3\ddot{p} =$$

$$= \int_{(\infty)} \bar{\Psi}^{1,2}(\vec{r}',\vec{v},t) \Psi^{1,2}(\vec{r}'',\vec{v},t) \delta(\vec{s}_1) d^3s_1 = \bar{\Psi}^{1,2}(\vec{r},\vec{v},t) \Psi^{1,2}(\vec{r},\vec{v},t),$$

where it is taken into account that

$$\delta(\vec{s}) = \frac{1}{(2\pi\hbar_2)^3} \int_{(\infty)} \exp\left(\pm i \frac{\vec{s}\vec{q}}{\hbar_2}\right) d^3q. \quad (A.8)$$

## Proof of Theorem 1

Let us substitute the representation of wave function $\Psi^{1,2}(\vec{r},\vec{v},t)$ (1.9) into the expression for Wigner function of the fourth rank (2.1), we obtain

$$W^{1,2,3,4} = \frac{1}{(2\pi\hbar_2)^6} \sum_{n,k} \rho_{kn} \int_{(\infty)}\int_{(\infty)} \overline{\psi}_n(\vec{r}',\vec{v}')\psi_k(\vec{r}'',\vec{v}'') e^{i\frac{\vec{s}_1\vec{\tilde{p}}-\vec{s}_2\vec{\tilde{p}}}{\hbar_2}} d^3s_1 d^3s_2. \quad (A.9)$$

Let us calculate the derivative $\partial_0 W^{1,2,3,4}$ of expression (A.9) taking into account (1.11)

$$(2\pi\hbar_2)^6 \partial_0 W^{1,2,3,4} = \sum_{n,k} \partial_0 \rho_{kn} \int_{(\infty)}\int_{(\infty)} \overline{\psi}_n(\vec{r}',\vec{v}')\psi_k(\vec{r}'',\vec{v}'') e^{i\frac{\vec{s}_1\vec{\tilde{p}}-\vec{s}_2\vec{\tilde{p}}}{\hbar_2}} d^3s_1 d^3s_2 =$$

$$= \frac{i}{\hbar_2} \sum_{n,k} \sum_j \int_{(\infty)}\int_{(\infty)} \psi_k(\vec{r}'',\vec{v}'') \rho_{kj} \overline{H}_{jn} \overline{\psi}_n(\vec{r}',\vec{v}') e^{i\frac{\vec{s}_1\vec{\tilde{p}}-\vec{s}_2\vec{\tilde{p}}}{\hbar_2}} d^3s_1 d^3s_2 -$$

$$- \frac{i}{\hbar_2} \sum_{n,k} \sum_j \int_{-\infty}^{+\infty}\int_{-\infty}^{+\infty} \psi_k(\vec{r}'',\vec{v}'') H_{kj} \rho_{jn} \overline{\psi}_n(\vec{r}',\vec{v}') e^{i\frac{\vec{s}_1\vec{\tilde{p}}-\vec{s}_2\vec{\tilde{p}}}{\hbar_2}} d^3s_1 d^3s_2 =$$

$$= \frac{i}{\hbar_2} \sum_{n,k} \overline{E}_n^{1,2} \sum_j \int_{(\infty)}\int_{(\infty)} \psi_k(\vec{r}'',\vec{v}'') \rho_{kj} \delta_{jn} \overline{\psi}_n(\vec{r}',\vec{v}') e^{i\frac{\vec{s}_1\vec{\tilde{p}}-\vec{s}_2\vec{\tilde{p}}}{\hbar_2}} d^3s_1 d^3s_2 -$$

$$- \frac{i}{\hbar_2} \sum_{n,k} E_k^{1,2} \sum_j \int_{(\infty)}\int_{(\infty)} \psi_k(\vec{r}'',\vec{v}'') \delta_{kj} \rho_{jn} \overline{\psi}_n(\vec{r}',\vec{v}') e^{i\frac{\vec{s}_1\vec{\tilde{p}}-\vec{s}_2\vec{\tilde{p}}}{\hbar_2}} d^3s_1 d^3s_2 =$$

$$= \frac{i}{\hbar_2} \sum_{n,k} \overline{E}_n^{1,2} \int_{(\infty)}\int_{(\infty)} \psi_k(\vec{r}'',\vec{v}'') \rho_{kn} \overline{\psi}_n(\vec{r}',\vec{v}') e^{i\frac{\vec{s}_1\vec{\tilde{p}}-\vec{s}_2\vec{\tilde{p}}}{\hbar_2}} d^3s_1 d^3s_2 -$$

$$- \frac{i}{\hbar_2} \sum_{n,k} E_k^{1,2} \int_{(\infty)}\int_{(\infty)} \psi_k(\vec{r}'',\vec{v}'') \rho_{kn} \overline{\psi}_n(\vec{r}',\vec{v}') e^{i\frac{\vec{s}_1\vec{\tilde{p}}-\vec{s}_2\vec{\tilde{p}}}{\hbar_2}} d^3s_1 d^3s_2,$$

$$\partial_0 W^{1,2,3,4} = \frac{i}{(2\pi\hbar_2)^6 \hbar_2} \sum_{n,k} \rho_{kn} \int_{(\infty)}\int_{(\infty)} \psi_k(\vec{r}'',\vec{v}'') \overline{\psi}_n(\vec{r}',\vec{v}') \left(\overline{E}_n^{1,2} - E_k^{1,2}\right) e^{i\frac{\vec{s}_1\vec{\tilde{p}}-\vec{s}_2\vec{\tilde{p}}}{\hbar_2}} d^3s_1 d^3s_2. \quad (A.10)$$

We define operators $\hat{H}'$ and $\hat{H}''$ as follows

$$\hat{H}' \stackrel{det}{=} -\frac{\hbar_2^2}{2m}\Delta_{v'} - i\hbar_2 \vec{v}'\nabla_{r'} + U^{1,2}(\vec{r}',\vec{v}'), \quad (A.11)$$

$$\hat{H}'' \stackrel{det}{=} -\frac{\hbar_2^2}{2m}\Delta_{v''} - i\hbar_2 \vec{v}''\nabla_{r''} + U^{1,2}(\vec{r}'',\vec{v}'').$$

Equations for wave functions $\psi_n(\vec{r}',\vec{v}')$ and $\psi_k(\vec{r}'',\vec{v}'')$ correspond to operators (A.11), i.e.

$$\hat{H}'\psi_n(\vec{r}',\vec{v}') = E_n^{1,2}\psi_n(\vec{r}',\vec{v}'), \tag{A.12}$$
$$\hat{H}''\psi_k(\vec{r}'',\vec{v}'') = E_k^{1,2}\psi_k(\vec{r}'',\vec{v}'').$$

Expressions (A.12) are present in integral (A.10), therefore

$$(2\pi\hbar_2)^6 \partial_0 W^{1,2,3,4} = \frac{i}{\hbar_2}\sum_{n,k}\rho_{kn}\int_{(\infty)}\int_{(\infty)}\psi_k(\vec{r}'',\vec{v}'')\overline{E}_n^{1,2}\overline{\psi}_n(\vec{r}',\vec{v}')e^{i\frac{\vec{s}_1\vec{p}-\vec{s}_2\vec{p}}{\hbar_2}}d^3s_1 d^3s_2 -$$

$$-\frac{i}{\hbar_2}\sum_{n,k}\rho_{kn}\int_{(\infty)}\int_{(\infty)}\overline{\psi}_n(\vec{r}',\vec{v}')E_k^{1,2}\psi_k(\vec{r}'',\vec{v}'')e^{i\frac{\vec{s}_1\vec{p}-\vec{s}_2\vec{p}}{\hbar_2}}d^3s_1 d^3s_2 = \tag{A.13}$$

$$=\frac{i}{\hbar_2}\sum_{n,k}\rho_{kn}\int_{(\infty)}\int_{(\infty)}\psi_k(\vec{r}'',\vec{v}'')\overline{\hat{H}}'\overline{\psi}_n(\vec{r}',\vec{v}')e^{i\frac{\vec{s}_1\vec{p}-\vec{s}_2\vec{p}}{\hbar_2}}d^3s_1 d^3s_2$$

$$-\frac{i}{\hbar_2}\sum_{n,k}\rho_{kn}\int_{(\infty)}\int_{(\infty)}\overline{\psi}_n(\vec{r}',\vec{v}')\hat{H}''\psi_k(\vec{r}'',\vec{v}'')e^{i\frac{\vec{s}_1\vec{p}-\vec{s}_2\vec{p}}{\hbar_2}}d^3s_1 d^3s_2,$$

where it is taken into account that $\overline{\hat{H}}'\overline{\psi}_n(\vec{r}',\vec{v}') = \overline{E}_n^{1,2}\overline{\psi}_n(\vec{r}',\vec{v}')$. We substitute expressions for operators (A.12) into equation (A.13), we obtain

$$(2\pi\hbar_2)^6 \partial_0 W^{1,2,3,4} =$$

$$=\frac{i}{\hbar_2}\sum_{n,k}\rho_{kn}\int_{(\infty)}\int_{(\infty)}\psi_k(\vec{r}'',\vec{v}'')\left[-\frac{\hbar_2^2}{2m}\Delta_{v'} - i\hbar_2\vec{v}'\nabla_{r'} + U^{1,2}(\vec{r}',\vec{v}')\right]\overline{\psi}_n(\vec{r}',\vec{v}')e^{i\frac{\vec{s}_1\vec{p}-\vec{s}_2\vec{p}}{\hbar_2}}d^3s_1 d^3s_2$$

$$-\frac{i}{\hbar_2}\sum_{n,k}\rho_{kn}\int_{(\infty)}\int_{(\infty)}\overline{\psi}_n(\vec{r}',\vec{v}')\left[-\frac{\hbar_2^2}{2m}\Delta_{v''} - i\hbar_2\vec{v}''\nabla_{r''} + U^{1,2}(\vec{r}'',\vec{v}'')\right]\psi_k(\vec{r}'',\vec{v}'')e^{i\frac{\vec{s}_1\vec{p}-\vec{s}_2\vec{p}}{\hbar_2}}d^3s_1 d^3s_2,$$

$$\partial_0 W^{1,2,3,4} = \mathcal{T} + \mathcal{U}, \tag{A.14}$$

where expressions $\mathcal{T}$ and $\mathcal{U}$ conditionally correspond to the kinetic and potential parts:

$$\mathcal{T} = -\frac{\hbar_2^2}{(2\pi\hbar_2)^6 2m}\frac{i}{\hbar_2}\sum_{n,k}\rho_{kn}\int_{(\infty)}\int_{(\infty)}e^{i\frac{\vec{s}_1\vec{p}-\vec{s}_2\vec{p}}{\hbar_2}}[\Delta_{v'} - \Delta_{v''}]\overline{\psi}_n(\vec{r}',\vec{v}')\psi_k(\vec{r}'',\vec{v}'')d^3s_1 d^3s_2 -$$

$$-\frac{1}{(2\pi\hbar_2)^6}\sum_{n,k}\rho_{kn}\int_{(\infty)}\int_{(\infty)}e^{i\frac{\vec{s}_1\vec{p}-\vec{s}_2\vec{p}}{\hbar_2}}[\vec{v}''\nabla_{r''} + \vec{v}'\nabla_{r'}]\overline{\psi}_n(\vec{r}',\vec{v}')\psi_k(\vec{r}'',\vec{v}'')d^3s_1 d^3s_2, \tag{A.15}$$

$$\mathcal{U} = \frac{i}{\hbar_2}\sum_{n,k}\rho_{kn}\times$$

$$\times\frac{1}{(2\pi\hbar_2)^6}\int_{(\infty)}\int_{(\infty)}e^{i\frac{\vec{s}_1\vec{p}-\vec{s}_2\vec{p}}{\hbar_2}}\left[U^{1,2}(\vec{r}',\vec{v}') - U^{1,2}(\vec{r}'',\vec{v}'')\right]\overline{\psi}_n(\vec{r}',\vec{v}')\psi_k(\vec{r}'',\vec{v}'')d^3s_1 d^3s_2. \tag{A.16}$$

Kinetic part $\mathcal{T}$ consists of two integrals (A.15). We transform the second integral into (A.15). Let us move from derivatives with respect to primed variables to derivatives with respect to non-primed variables:

$$\vec{r} = \frac{\vec{r}' + \vec{r}''}{2}, \ \vec{s}_1 = \vec{r}'' - \vec{r}', \ \vec{v} = \frac{\vec{v}' + \vec{v}''}{2}, \ \vec{s}_2 = \vec{v}'' - \vec{v}'. \tag{A.17}$$

$$\nabla_{r'} = \frac{1}{2}\nabla_r - \nabla_{s_1}, \qquad \nabla_{r''} = \frac{1}{2}\nabla_r + \nabla_{s_1},$$

$$\vec{v}'\nabla_{r'} + \vec{v}''\nabla_{r''} = \left(\vec{v} - \frac{\vec{s}_2}{2}\right)\left(\frac{1}{2}\nabla_r - \nabla_{s_1}\right) + \left(\vec{v} + \frac{\vec{s}_2}{2}\right)\left(\frac{1}{2}\nabla_r + \nabla_{s_1}\right) = \vec{v}\nabla_r + \vec{s}_2\nabla_{s_1}, \tag{A.18}$$

$$\frac{1}{(2\pi\hbar_2)^6}\sum_{n,k}\rho_{kn}\int_{(\infty)}\int_{(\infty)} e^{i\frac{\vec{s}_1\vec{\dot{p}} - \vec{s}_2\vec{\dot{p}}}{\hbar_2}}\left[\vec{v}''\nabla_{r''} + \vec{v}'\nabla_{r'}\right]\overline{\psi}_n(\vec{r}',\vec{v}')\psi_k(\vec{r}'',\vec{v}'')d^3s_1 d^3s_2 =$$

$$= \vec{v}\nabla_r \sum_{n,k}\rho_{kn}\frac{1}{(2\pi\hbar_2)^6}\int_{(\infty)}\int_{(\infty)} e^{i\frac{\vec{s}_1\vec{\dot{p}} - \vec{s}_2\vec{\dot{p}}}{\hbar_2}}\overline{\psi}_n(\vec{r}',\vec{v}')\psi_k(\vec{r}'',\vec{v}'')d^3s_1 d^3s_2 +$$

$$+ \sum_{n,k}\rho_{kn}\frac{1}{(2\pi\hbar_2)^6}\int_{(\infty)}\nabla_{s_1}\left[\overline{\psi}_n(\vec{r}',\vec{v}')\psi_k(\vec{r}'',\vec{v}'')\right]e^{i\frac{\vec{s}_1\vec{\dot{p}}}{\hbar_2}}d^3s_1 \int_{(\infty)} e^{-i\frac{\vec{s}_2\vec{\dot{p}}}{\hbar_2}}\vec{s}_2 d^3s_2 =$$

$$= \vec{v}\nabla_r W^{1,2,3,4} - i\frac{\vec{\dot{p}}}{\hbar_2}\sum_{n,k}\rho_{kn}\frac{1}{(2\pi\hbar_2)^6}\int_{(\infty)}\left[\overline{\psi}_n(\vec{r}',\vec{v}')\psi_k(\vec{r}'',\vec{v}'')\right]e^{i\frac{\vec{s}_1\vec{\dot{p}}}{\hbar_2}}d^3s_1 \int_{(\infty)} e^{-i\frac{\vec{s}_2\vec{\dot{p}}}{\hbar_2}}\vec{s}_2 d^3s_2 =$$

$$= \vec{v}\nabla_r W^{1,2,3,4} - i\frac{\vec{\dot{p}}}{\hbar_2}i\hbar_2\nabla_{\dot{p}}\sum_{n,k}\rho_{kn}\frac{1}{(2\pi\hbar_2)^6}\int_{(\infty)}\overline{\psi}_n(\vec{r}',\vec{v}')\psi_k(\vec{r}'',\vec{v}'')e^{i\frac{\vec{s}_1\vec{\dot{p}}}{\hbar_2}}d^3s_1 \int_{(\infty)} e^{-i\frac{\vec{s}_2\vec{\dot{p}}}{\hbar_2}}d^3s_2 =$$

$$= \vec{v}\nabla_r W^{1,2,3,4} + \vec{\dot{p}}\nabla_{\dot{p}}\sum_{n,k}\rho_{kn}\frac{1}{(2\pi\hbar_2)^6}\int_{(\infty)}\int_{(\infty)} e^{i\frac{\vec{s}_1\vec{\dot{p}} - \vec{s}_2\vec{\dot{p}}}{\hbar_2}}\overline{\psi}_n(\vec{r}',\vec{v}')\psi_k(\vec{r}'',\vec{v}'')d^3s_1 d^3s_2,$$

$$\frac{1}{(2\pi\hbar_2)^6}\sum_{n,k}\rho_{kn}\int_{(\infty)}\int_{(\infty)} e^{i\frac{\vec{s}_1\vec{\dot{p}} - \vec{s}_2\vec{\dot{p}}}{\hbar_2}}\left[\vec{v}''\nabla_{r''} + \vec{v}'\nabla_{r'}\right]\overline{\psi}_n(\vec{r}',\vec{v}')\psi_k(\vec{r}'',\vec{v}'')d^3s_1 d^3s_2 = \tag{A.19}$$

$$= \vec{v}\nabla_r W^{1,2,3,4} + \vec{\dot{p}}\nabla_{\dot{p}} W^{1,2,3,4},$$

where it is taken into account that $e^{-i\frac{\vec{s}_2\vec{\dot{p}}}{\hbar_2}}\vec{s}_2 = i\hbar_2\nabla_{\dot{p}} e^{-i\frac{\vec{s}_2\vec{\dot{p}}}{\hbar_2}}$.

We consider the first integral in the kinetic part $\mathcal{T}$ of (A.15). Let us make the transition to non-primed coordinates (A.17):

$$\nabla_{v'} = \frac{1}{2}\nabla_v - \nabla_{s_2}, \qquad \nabla_{v''} = \frac{1}{2}\nabla_v + \nabla_{s_2}$$

$$\Delta_{v'} = (\nabla_{v'},\nabla_{v'}) = \left(\frac{1}{2}\nabla_v - \nabla_{s_2}, \frac{1}{2}\nabla_v - \nabla_{s_2}\right) = \frac{1}{4}\Delta_v - (\nabla_v,\nabla_{s_2}) + \Delta_{s_2},$$

$$\Delta_{v''} = \frac{1}{4}\Delta_v + (\nabla_v,\nabla_{s_2}) + \Delta_{s_2},$$

$$\Delta_{v'} - \Delta_{v''} = -2(\nabla_v,\nabla_{s_2}). \tag{A.20}$$

Let us substitute expression (A.20) into the first integral (A.15), we obtain

$$-\frac{i\hbar_2}{2m}\frac{1}{(2\pi\hbar_2)^6}\sum_{n,k}\rho_{kn}\int\limits_{(\infty)}\int\limits_{(\infty)}e^{i\frac{\vec{s}_1\dot{\vec{p}}-\vec{s}_2\dot{\vec{p}}}{\hbar_2}}[\Delta_{\vec{v}'}-\Delta_{\vec{v}''}]\overline{\psi}_n(\vec{r}',\vec{v}')\psi_k(\vec{r}'',\vec{v}'')d^3s_1d^3s_2 =$$

$$=\frac{i\hbar_2}{m}\frac{1}{(2\pi\hbar_2)^6}\left(\nabla_v,\sum_{n,k}\rho_{kn}\int\limits_{(\infty)}e^{-i\frac{\vec{s}_2\dot{\vec{p}}}{\hbar_2}}\nabla_{s_2}\left[\overline{\psi}_n(\vec{r}',\vec{v}')\psi_k(\vec{r}'',\vec{v}'')\right]d^3s_2\int\limits_{(\infty)}e^{i\frac{\vec{s}_1\dot{\vec{p}}}{\hbar_2}}d^3s_1\right)=$$

$$=i\frac{\dot{\vec{p}}}{\hbar_2}\frac{i\hbar_2}{m}\frac{1}{(2\pi\hbar_2)^6}\left(\nabla_v,\sum_{n,k}\rho_{kn}\int\limits_{(\infty)}\overline{\psi}_n(\vec{r}',\vec{v}')\psi_k(\vec{r}'',\vec{v}'')e^{-i\frac{\vec{s}_2\dot{\vec{p}}}{\hbar_2}}d^3s_2\int\limits_{(\infty)}e^{i\frac{\vec{s}_1\dot{\vec{p}}}{\hbar_2}}d^3s_1\right)=$$

$$=-\left(\frac{\dot{\vec{p}}}{m},\nabla_v\right)\frac{1}{(2\pi\hbar_2)^6}\sum_{n,k}\rho_{kn}\int\limits_{(\infty)}\int\limits_{(\infty)}\overline{\psi}_n(\vec{r}',\vec{v}')\psi_k(\vec{r}'',\vec{v}'')e^{i\frac{\vec{s}_1\ddot{\vec{p}}-\vec{s}_2\dot{\vec{p}}}{\hbar_2}}d^3s_1d^3s_2,$$

$$-\frac{i\hbar_2}{2m}\frac{1}{(2\pi\hbar_2)^6}\sum_{n,k}\rho_{kn}\int\limits_{(\infty)}\int\limits_{(\infty)}e^{i\frac{\vec{s}_1\ddot{\vec{p}}-\vec{s}_2\dot{\vec{p}}}{\hbar_2}}[\Delta_{\vec{v}'}-\Delta_{\vec{v}''}]\overline{\psi}_n(\vec{r}',\vec{v}')\psi_k(\vec{r}'',\vec{v}'')d^3s_1d^3s_2 = -\frac{\dot{\vec{p}}}{m}\nabla_v W^{1,2,3,4}. \quad (A.21)$$

Substituting (A.19) and (A.21) into expression (A.15), we obtain

$$\mathcal{T}=-\left(\frac{\vec{p}}{m}\nabla_r+\dot{\vec{p}}\nabla_p+\ddot{\vec{p}}\nabla_{\dot{p}}\right)W^{1,2,3,4}. \quad (A.22)$$

Let us transform the potential part (A.16). By the condition of the theorem, potential $U^{1,2}(\vec{r},\vec{v})$ admits expansion in a Taylor series:

$$U^{1,2}\left(\vec{r}-\frac{\vec{s}_1}{2},\vec{v}-\frac{\vec{s}_2}{2}\right)=\sum_{n=0}^{+\infty}\sum_{k=0}^{+\infty}\frac{(-1)^{n+k}}{n!k!}\left(\frac{\vec{s}_1}{2}\nabla_r\right)^n\left(\frac{\vec{s}_2}{2}\nabla_v\right)^k U^{1,2}(\vec{r},\vec{v}), \quad (A.23)$$

$$U^{1,2}\left(\vec{r}+\frac{\vec{s}_1}{2},\vec{v}+\frac{\vec{s}_2}{2}\right)=\sum_{n=0}^{+\infty}\sum_{k=0}^{+\infty}\frac{1}{n!k!}\left(\frac{\vec{s}_1}{2}\nabla_r\right)^n\left(\frac{\vec{s}_2}{2}\nabla_v\right)^k U^{1,2}(\vec{r},\vec{v}).$$

The form of writing (A.23) for the Taylor series is derived from the exponential expansion of the differential operator. For example, for function $U(\vec{r})$ of three variables, the Taylor series is as follows

$$U(\vec{r}+\Delta\vec{r})=\sum_{n_1=0}^{+\infty}\sum_{n_2=0}^{+\infty}\sum_{n_3=0}^{+\infty}\frac{\Delta x^{n_1}\Delta y^{n_2}\Delta z^{n_3}}{n_1!n_2!n_3!}\frac{\partial^{n_1+n_2+n_3}U(\vec{r})}{\partial x^{n_1}\partial y^{n_2}\partial z^{n_3}}=$$

$$=\sum_{n_1=0}^{+\infty}\frac{1}{n_1!}\left(\Delta x\frac{\partial}{\partial x}\right)^{n_1}\sum_{n_2=0}^{+\infty}\frac{1}{n_2!}\left(\Delta y\frac{\partial}{\partial y}\right)^{n_2}\sum_{n_3=0}^{+\infty}\frac{1}{n_3!}\left(\Delta z\frac{\partial}{\partial z}\right)^{n_3}U(\vec{r})=$$

$$=e^{\Delta x\frac{\partial}{\partial x}}e^{\Delta y\frac{\partial}{\partial y}}e^{\Delta z\frac{\partial}{\partial z}}U(\vec{r})=e^{\Delta x\frac{\partial}{\partial x}+\Delta y\frac{\partial}{\partial y}+\Delta z\frac{\partial}{\partial z}}U(\vec{r})=\exp(\Delta\vec{r},\nabla_r)U(\vec{r})=\sum_{n=0}^{+\infty}\frac{(\Delta\vec{r},\nabla_r)^n}{n!}U(\vec{r}),$$

where it is taken into consideration that, due to the analyticity of function $U(\vec{r})$, its mixed derivatives are equal and do not commute in the exponent. Thus expressions (A.23) are correct, therefore

$$U^{1,2}(\vec{r}'',\vec{v}'') - U^{1,2}(\vec{r}',\vec{v}') = \sum_{n=0}^{+\infty}\sum_{k=0}^{+\infty} \frac{1-(-1)^{n+k}}{n!k!} \left(\frac{\vec{s}_1}{2}\nabla_r\right)^n \left(\frac{\vec{s}_2}{2}\nabla_v\right)^k U^{1,2}(\vec{r},\vec{v}) =$$

$$= \sum_{l=0}^{+\infty}\sum_{n=0}^{+\infty} \frac{(\vec{s}_1\nabla_r)^n (\vec{s}_2\nabla_v)^{2l-n+1}}{2^{2l} n!(2l-n+1)!} U^{1,2}(\vec{r},\vec{v}) =$$

$$= i\hbar_2 \sum_{l=0}^{+\infty}\sum_{n=0}^{+\infty} \frac{(-1)^n (i\hbar_2/2)^{2l}}{n!(2l-n+1)!} \left(i\frac{\vec{s}_1}{\hbar_2}\nabla_r\right)^n \left(-i\frac{\vec{s}_2}{\hbar_2}\nabla_v\right)^{2l-n+1} U^{1,2}(\vec{r},\vec{v}),$$

$$U^{1,2}(\vec{r}'',\vec{v}'') - U^{1,2}(\vec{r}',\vec{v}') = i\hbar_2 \sum_{l=0}^{+\infty}\sum_{n=0}^{+\infty} \frac{(-1)^{n+l}(\hbar_2/2)^{2l}}{n!(2l-n+1)!} \left(i\frac{\vec{s}_1}{\hbar_2}\nabla_r\right)^n \left(-i\frac{\vec{s}_2}{\hbar_2}\nabla_v\right)^{2l-n+1} U^{1,2}(\vec{r},\vec{v}).$$

(A.24)

The following relation is true

$$e^{i\frac{\vec{s}_1\vec{\ddot{p}}}{\hbar_2}} \left(i\frac{\vec{s}_1}{\hbar_2}\nabla_r\right)^n e^{-i\frac{\vec{s}_2\vec{\dot{p}}}{\hbar_2}} \left(-i\frac{\vec{s}_2}{\hbar_2}\nabla_v\right)^{2l-n+1} U^{1,2}(\vec{r},\vec{v}) = \left(\nabla_{\ddot{p}},\nabla_r\right)^n \left(\nabla_{\dot{p}},\nabla_v\right)^{2l-n+1} e^{i\frac{\vec{s}_1\vec{\ddot{p}}-\vec{s}_2\vec{\dot{p}}}{\hbar_2}} U^{1,2}(\vec{r},\vec{v}).$$

(A.25)

Taking into account expressions (A.24), (A.25), the potential part (A.16) takes the form:

$$\mathcal{U} = -\frac{i}{\hbar_2} i\hbar_2 \sum_{l=0}^{+\infty}\sum_{n=0}^{+\infty} \frac{(-1)^{n+l}(\hbar_2/2)^{2l}}{n!(2l-n+1)!} U^{1,2}(\vec{r},\vec{v}) \left(\overleftarrow{\nabla}_r,\vec{\nabla}_{\ddot{p}}\right)^n \left(\overleftarrow{\nabla}_v,\vec{\nabla}_{\dot{p}}\right)^{2l-n+1} \times$$

$$\times \frac{1}{(2\pi\hbar_2)^3} \sum_{n,k} \rho_{kn} \int_{(\infty)}\int_{(\infty)} e^{i\frac{\vec{s}_1\vec{\ddot{p}}-\vec{s}_2\vec{\dot{p}}}{\hbar_2}} \overline{\psi}_n(\vec{r}',\vec{v}')\psi_k(\vec{r}'',\vec{v}'') d^3s_1 d^3s_2,$$

$$\mathcal{U} = \sum_{l=0}^{+\infty}\sum_{n=0}^{+\infty} \frac{(-1)^{n+l}(\hbar_2/2)^{2l}}{n!(2l-n+1)!} U^{1,2}(\vec{r},\vec{v}) \left(\overleftarrow{\nabla}_r,\vec{\nabla}_{\ddot{p}}\right)^n \left(\overleftarrow{\nabla}_v,\vec{\nabla}_{\dot{p}}\right)^{2l-n+1} W^{1,2,3,4}. \quad (A.26)$$

Substituting the kinetic (A.22) and potential (A.26) parts into the initial equation (A.14), we obtain

$$\left(\partial_0 + \frac{\vec{p}}{m}\nabla_r + \vec{\dot{p}}\nabla_p + \vec{\ddot{p}}\nabla_{\dot{p}}\right) W^{1,2,3,4} =$$

$$= \sum_{l=0}^{+\infty}\sum_{n=0}^{+\infty} \frac{(-1)^{n+l}(\hbar_2/2)^{2l}}{n!(2l-n+1)!} U^{1,2}(\vec{r},\vec{v}) \left(\overleftarrow{\nabla}_r,\vec{\nabla}_{\ddot{p}}\right)^n \left(\overleftarrow{\nabla}_v,\vec{\nabla}_{\dot{p}}\right)^{2l-n+1} W^{1,2,3,4}.$$

(A.27)

Theorem 1 is proved.

*Proof of Theorem 2*

Taking into account the independence of kinematical values $\vec{r},\vec{v},\vec{\dot{v}},\vec{\ddot{v}}$, we will directly substitute expression (2.9) into equation (2.10), we obtain

$$\frac{\partial f^{1,2,3,4}}{\partial t}+v_\lambda\frac{\partial f^{1,2,3,4}}{\partial x_\lambda}+\dot v_\lambda\frac{\partial f^{1,2,3,4}}{\partial v_\lambda}+\ddot v_\lambda\frac{\partial f^{1,2,3,4}}{\partial \dot v_\lambda}+\frac{\partial}{\partial \ddot v_\mu}\left[\frac{1}{m}\sum_{l=0}^{+\infty}\frac{(-1)^l(\hbar_2/2m)^{2l}}{(2l+1)!}\frac{\partial^{2l+1}U_1}{\partial x_\mu^{2l+1}}\frac{\partial^{2l}f^{1,2,3,4}}{\partial \ddot v_\mu^{2l}}\right]=0,$$

$$\left[\frac{\partial}{\partial t}+v_\lambda\frac{\partial}{\partial x_\lambda}+\dot v_\lambda\frac{\partial}{\partial v_\lambda}+\ddot v_\lambda\frac{\partial}{\partial \dot v_\lambda}\right]f^{1,2,3,4}+\frac{1}{m}\frac{\partial U_1}{\partial x_\mu}\frac{\partial f^{1,2,3,4}}{\partial \ddot v_\mu}+$$

$$+\frac{1}{m}\sum_{l=1}^{+\infty}\frac{(-1)^l(\hbar_2/2m)^{2l}}{(2l+1)!}\frac{\partial^{2l+1}U_1}{\partial x_\mu^{2l+1}}\frac{\partial^{2l+1}f^{1,2,3,4}}{\partial \ddot v_\mu^{2l+1}}=0,$$

(A.28)

which coincides with equation (2.8). Let us find the mean kinematical values:

$$f^{1,2,3}\langle\ddot v_\mu\rangle_{1,2,3}=\int_{(\infty)}f^{1,2,3,4}\langle\ddot v_\mu\rangle_{1,2,3,4}d^3\ddot v=\frac{1}{m}\sum_{l=0}^{+\infty}\frac{(-1)^l(\hbar_2/2m)^{2l}}{(2l+1)!}\frac{\partial^{2l+1}U_1}{\partial x_\mu^{2l+1}}\int_{(\infty)}\frac{\partial^{2l}f^{1,2,3,4}}{\partial \ddot v_\mu^{2l}}d^3\ddot v=$$

$$=\frac{1}{m}\frac{\partial U_1}{\partial x_\mu}\int_{(\infty)}f^{1,2,3,4}d^3\ddot v=\frac{1}{m}\frac{\partial U_1}{\partial x_\mu}f^{1,2,3},$$

(A.29)

$$f^{1,2}\langle\ddot v_\mu\rangle_{1,2}=\int_{(\infty)}f^{1,2,3}\langle\ddot v_\mu\rangle_{1,2,3}d^3\dot v=\frac{1}{m}\frac{\partial U_1}{\partial x_\mu}f^{1,2},$$

(A.30)

where the Vlasov condition of fast tending to zero of distribution functions at infinity is taken into account [18, 19]. Theorem 2 is proved.

*Proof of Theorem 3*
Let us write equation (2.7) in the form:

$$\frac{\partial W^{1,2,3,4}}{\partial t}+\operatorname{div}_r\left[W^{1,2,3,4}\vec v\right]+\operatorname{div}_v\left[W^{1,2,3,4}\dot{\vec v}\right]+\operatorname{div}_{\dot v}\left[W^{1,2,3,4}\left(\ddot{\vec v}-\frac{1}{m}\nabla_v U^{1,2}\right)\right]+$$

$$+\operatorname{div}_{\ddot v}\left[W^{1,2,3,4}\frac{1}{m}\nabla_r U^{1,2}\right]=\frac{1}{m}\sum_{n=0}^{+\infty}\sum_{l=1}^{+\infty}\frac{(-1)^{n+l}(\hbar_2/2m)^{2l}}{n!(2l-n+1)!}U^{1,2}\left(\vec\nabla_r,\vec\nabla_{\ddot v}\right)^n\left(\vec\nabla_v,\vec\nabla_{\dot v}\right)^{2l-n+1}W^{1,2,3,4},$$

(A.31)

We integrate equation (A.31) over space $\int_{(\infty)}d^3\ddot v$, we obtain:

$$\frac{\partial W^{1,2,3}}{\partial t}+\operatorname{div}_r\left[W^{1,2,3}\vec v\right]+\operatorname{div}_v\left[W^{1,2,3}\dot{\vec v}\right]+\operatorname{div}_{\dot v}\left[W^{1,2,3}\langle\ddot{\vec v}\rangle_{1,2,3}\right]-\frac{1}{m}\operatorname{div}_{\dot v}\left[W^{1,2,3}\nabla_v U^{1,2}\right]=$$

$$=\frac{1}{m}\sum_{n=0}^{+\infty}\sum_{l=1}^{+\infty}\frac{(-1)^{n+l}(\hbar_2/2m)^{2l}}{n!(2l-n+1)!}\frac{\partial^{2l+1}U^{1,2}}{\partial x_\lambda^n\partial v_\mu^{2l-n+1}}\frac{\vec\partial^{2l-n+1}}{\partial\dot v_\mu^{2l-n+1}}\int\frac{\vec\partial^n}{\partial\ddot v_\lambda^n}W^{1,2,3,4}d\ddot v=$$

$$=\frac{1}{m}\sum_{l=1}^{+\infty}\frac{(-1)^l(\hbar_2/2m)^{2l}}{(2l+1)!}\frac{\partial^{2l+1}U^{1,2}}{\partial v_\mu^{2l+1}}\frac{\partial^{2l+1}W^{1,2,3}}{\partial\dot v_\mu^{2l+1}},$$

(A.32)

where the Ostrogradsky-Gauss theorem and the Vlasov condition of fast tending to zero of distribution function at infinity are used [18, 19]. As a result, from (A.32) it follows that

$$\frac{\partial W^{1,2,3}}{\partial t} + \text{div}_r \left[ W^{1,2,3} \vec{v} \right] + \text{div}_v \left[ W^{1,2,3} \vec{\dot{v}} \right] + \text{div}_{\dot{v}} \left[ W^{1,2,3} \langle \vec{\ddot{v}} \rangle_{1,2,3} \right] =$$

$$= \frac{1}{m} \frac{\partial U^{1,2}}{\partial v_\mu} \frac{\partial W^{1,2,3}}{\partial \dot{v}_\mu} + \frac{1}{m} \sum_{l=1}^{+\infty} \frac{(-1)^l (\hbar_2/2m)^{2l}}{(2l+1)!} \frac{\partial^{2l+1} U^{1,2}}{\partial v_\mu^{2l+1}} \frac{\partial^{2l+1} W^{1,2,3}}{\partial \dot{v}_\mu^{2l+1}} = \quad (A.33)$$

$$= \frac{1}{m} \sum_{l=0}^{+\infty} \frac{(-1)^l (\hbar_2/2m)^{2l}}{(2l+1)!} \frac{\partial^{2l+1} U^{1,2}}{\partial v_\mu^{2l+1}} \frac{\partial^{2l+1} W^{1,2,3}}{\partial \dot{v}_\mu^{2l+1}},$$

Resulting equation (A.33) coincides with equation (2.12). In a similar way, we integrate equation (A.31) over space $\int_{(\infty)} d^3 \dot{v}$, we obtain

$$\frac{\partial W^{1,2,4}}{\partial t} + \text{div}_r \left[ W^{1,2,4} \vec{v} \right] + \text{div}_v \left[ W^{1,2,4} \langle \vec{\dot{v}} \rangle_{1,2,4} \right] + \text{div}_{\ddot{v}} \left[ W^{1,2,4} \frac{1}{m} \nabla_r U^{1,2} \right] =$$

$$= \frac{1}{m} \sum_{n=0}^{+\infty} \sum_{l=1}^{+\infty} \frac{(-1)^{n+l} (\hbar_2/2m)^{2l}}{n!(2l-n+1)!} \frac{\partial^{2l+1} U^{1,2}}{\partial x_\lambda^n \partial v_\mu^{2l-n+1}} \frac{\vec{\partial}^n}{\partial \ddot{v}_\lambda^n} \int \frac{\vec{\partial}^{2l-n+1}}{\partial \dot{v}_\mu^{2l-n+1}} W^{1,2,3,4} d^3 \dot{v} =$$

$$= \frac{1}{m} \sum_{l=1}^{+\infty} \frac{(-1)^{l+1} (\hbar_2/2m)^{2l}}{(2l+1)!} \frac{\partial^{2l+1} U^{1,2}}{\partial x_\lambda^{2l+1}} \frac{\partial^{2l+1} W^{1,2,4}}{\partial \ddot{v}_\lambda^{2l+1}},$$

from here

$$\frac{\partial W^{1,2,4}}{\partial t} + \text{div}_r \left[ W^{1,2,4} \vec{v} \right] + \text{div}_v \left[ W^{1,2,4} \langle \vec{\dot{v}} \rangle_{1,2,4} \right] =$$

$$= -\frac{1}{m} \frac{\partial U^{1,2}}{\partial x_\lambda} \frac{\vec{\partial} W^{1,2,4}}{\partial \ddot{v}_\lambda} + \frac{1}{m} \sum_{l=1}^{+\infty} \frac{(-1)^{l+1} (\hbar_2/2m)^{2l}}{(2l+1)!} \frac{\partial^{2l+1} U^{1,2}}{\partial x_\lambda^{2l+1}} \frac{\vec{\partial}^{2l+1} W^{1,2,4}}{\partial \ddot{v}_\lambda^{2l+1}} = \quad (A.34)$$

$$= \frac{1}{m} \sum_{l=0}^{+\infty} \frac{(-1)^{l+1} (\hbar_2/2m)^{2l}}{(2l+1)!} \frac{\partial^{2l+1} U^{1,2}}{\partial x_\lambda^{2l+1}} \frac{\vec{\partial}^{2l+1} W^{1,2,4}}{\partial \ddot{v}_\lambda^{2l+1}}.$$

Next, we use the second Vlasov-Moyal approximation (2.9) with potential $U^{1,2}$, integrating it over the acceleration space:

$$W^{1,2,4} \langle \ddot{v}_\mu \rangle_{1,2,4} = \frac{1}{m} \sum_{l=0}^{+\infty} \frac{(-1)^l (\hbar_2/2m)^{2l}}{(2l+1)!} \frac{\partial^{2l+1} U^{1,2}}{\partial x_\mu^{2l+1}} \frac{\partial^{2l} W^{1,2,4}}{\partial \ddot{v}_\mu^{2l}},$$

$$\text{div}_{\ddot{v}} \left[ W^{1,2,4} \langle \vec{\ddot{v}} \rangle_{1,2,4} \right] = -\sum_{l=0}^{+\infty} \frac{(-1)^{l+1} (\hbar_2/2m)^{2l}}{m(2l+1)!} U^{1,2} \left( \overleftarrow{\nabla}_r, \vec{\nabla}_{\ddot{v}} \right)^{2l+1} W^{1,2,4}. \quad (A.35)$$

Expression (A.35) coincides with the right-hand side of equation (A.34). Therefore, equation (A.34) coincides with equation (2.13).

Equation (2.14) is obtained by integrating either equation (2.13) over space $\int_{(\infty)} d^3 \ddot{v}$, or by integrating equation (2.12) over space $\int_{(\infty)} d^3 \dot{v}$. In both cases, the same equation (2.14) will be obtained. Let us integrate equation (A.24):

$$\frac{\partial W^{1,2}}{\partial t} + \text{div}_r\left[W^{1,2}\vec{v}\right] + \text{div}_v\left[W^{1,2}\langle\dot{\vec{v}}\rangle_{1,2}\right] =$$
$$= \frac{1}{m}\sum_{l=0}^{+\infty}\frac{(-1)^{l+1}(\hbar_2/2m)^{2l}}{(2l+1)!}\frac{\partial^{2l+1}U^{1,2}}{\partial x_\lambda^{2l+1}}\int_{(\infty)}\frac{\partial^{2l+1}W^{1,2,4}}{\partial \dot{v}_\lambda^{2l+1}}d^3\ddot{v} = 0. \tag{A.35}$$

Expression (A.35) proves equation (2.14) is true. The Theorem is proved.

## Appendix B

Let us calculate Wigner function $W^{1,2,3,4}$ using expressions (3.1), (2.1) and (3.4), we obtain:

$$\Psi^{1,2}(x,v,t) = \sqrt{\frac{m}{\pi\hbar}}\cdot\exp\left[-\frac{1}{\hbar\omega}\left(\frac{mv^2}{2}+\frac{m\omega^2 x^2}{2}\right)-i\left(\frac{m\omega^2}{\hbar_2}xv+\frac{E^{1,2}}{\hbar_2}t\right)\right],$$

$$W^{1,2,3,4} = \frac{1}{(2\pi\hbar_2)^2}\frac{m}{\pi\hbar}\int_{-\infty}^{+\infty}\int_{-\infty}^{+\infty}e^{-\frac{m}{2\hbar\omega}\left[\left(v-\frac{s_2}{2}\right)^2+\left(v+\frac{s_2}{2}\right)^2+\omega^2\left(x-\frac{s_1}{2}\right)^2+\omega^2\left(x+\frac{s_1}{2}\right)^2\right]}e^{-i\frac{m}{\hbar}\left[\left(x+\frac{s_1}{2}\right)\left(v+\frac{s_2}{2}\right)-\left(x-\frac{s_1}{2}\right)\left(v-\frac{s_2}{2}\right)\right]}e^{i\frac{s_1\dot{p}-s_2\dot{p}}{\hbar_2}}ds_1 ds_2 =$$

$$= \frac{1}{(2\pi\hbar_2)^2}\frac{m}{\pi\hbar}\int_{-\infty}^{+\infty}\int_{-\infty}^{+\infty}e^{-\frac{m}{2\hbar\omega}\left[2v^2+\frac{s_2^2}{2}+\omega^2\left(2x^2+\frac{s_1^2}{2}\right)\right]}e^{-i\frac{m}{\hbar}(xs_2+vs_1)}e^{i\frac{s_1\ddot{p}-s_2\dot{p}}{\hbar_2}}ds_1 ds_2 =$$

$$= \frac{1}{(2\pi\hbar_2)^2}\frac{m}{\pi\hbar}e^{-\frac{m}{\hbar\omega}(v^2+\omega^2 x^2)}\int_{-\infty}^{+\infty}\int_{-\infty}^{+\infty}e^{-\frac{m}{4\hbar\omega}(s_2^2+\omega^2 s_1^2)}e^{-i\frac{m}{\hbar}(xs_2+vs_1)}e^{i\frac{s_1\ddot{p}-s_2\dot{p}}{\hbar_2}}ds_1 ds_2 =$$

$$= \frac{1}{(2\pi\hbar_2)^2}\frac{m}{\pi\hbar}e^{-\frac{m}{\hbar\omega}(v^2+\omega^2 x^2)}\int_{-\infty}^{+\infty}e^{-\frac{m\omega}{4\hbar}s_1^2}e^{-i\frac{mv}{\hbar}s_1}e^{i\frac{\ddot{p}}{\hbar_2}s_1}ds_1\int_{-\infty}^{+\infty}e^{-\frac{m}{4\hbar\omega}s_2^2}e^{-i\frac{mx}{\hbar}s_2}e^{-i\frac{\dot{p}}{\hbar_2}s_2}ds_2 =$$

$$W^{1,2,3,4} = \frac{1}{(2\pi\hbar_2)^2}\frac{m}{\pi\hbar}e^{-\frac{m}{\hbar\omega}(v^2+\omega^2 x^2)}\int_{-\infty}^{+\infty}e^{-\frac{m\omega}{4\hbar}\left[s_1^2+i\frac{4}{\omega^3}(\omega^2 v-\ddot{v})s_1\right]}ds_1\int_{-\infty}^{+\infty}e^{-\frac{m}{4\hbar\omega}\left[s_2^2+i\frac{4}{\omega}(\omega^2 x+\dot{v})s_2\right]}ds_2. \tag{B.1}$$

We transform the expressions in the exponent:

$$-\frac{m\omega}{4\hbar}\left[s_1^2+i\frac{4}{\omega^3}(\omega^2 v-\ddot{v})s_1\right] = -\frac{m\omega}{4\hbar}\left[s_1+\frac{2i}{\omega^3}(\omega^2 v-\ddot{v})\right]^2-\frac{m}{\hbar\omega^5}(\omega^2 v-\ddot{v})^2, \tag{B.2}$$

$$-\frac{m}{4\hbar\omega}\left[s_2^2+i\frac{4}{\omega}(\omega^2 x+\dot{v})s_2\right] = -\frac{m}{4\hbar\omega}\left[s_2+\frac{2i}{\omega}(\omega^2 x+\dot{v})\right]^2-\frac{m}{\hbar\omega^3}(\omega^2 x+\dot{v})^2.$$

Let us introduce the notations:

$$z_1 = \frac{1}{2}\sqrt{\frac{m\omega}{\hbar}}\left[s_1+\frac{2i}{\omega^3}(\omega^2 v-\ddot{v})\right], \qquad z_2 = \frac{1}{2}\sqrt{\frac{m}{\hbar\omega}}\left[s_2+\frac{2i}{\omega}(\omega^2 x+\dot{v})\right]. \tag{B.3}$$

Using expressions (B.2) and notations (B.3), integral (B.1) takes the form:

$$W^{1,2,3,4} = \frac{1}{(2\pi\hbar_2)^2}\frac{m}{\pi\hbar} e^{-\frac{m}{\hbar\omega}(v^2+\omega^2 x^2)-\frac{m}{\hbar\omega^5}(\omega^2 v-\dot{v})^2-\frac{m}{\hbar\omega^3}(\omega^2 x+\dot{v})^2} \int_{-\infty}^{+\infty} e^{-z_1^2} ds_1 \int_{-\infty}^{+\infty} e^{-z_2^2} ds_2 =$$

$$= \frac{1}{(2\pi\hbar_2)^2}\frac{m}{\pi\hbar}\exp\left[-\frac{m}{\hbar\omega}(v^2+\omega^2 x^2)-\frac{m}{\hbar\omega^5}(\omega^2 v-\ddot{v})^2-\frac{m}{\hbar\omega^3}(\omega^2 x+\dot{v})^2\right]4\pi\sqrt{\frac{\hbar\omega}{m}}\sqrt{\frac{\hbar}{m\omega}},$$

$$W^{1,2,3,4} = \frac{1}{(\pi\hbar_2)^2}\exp\left[-\frac{m}{\hbar\omega}(v^2+\omega^2 x^2)-\frac{m}{\hbar\omega^5}(\omega^2 v-\ddot{v})^2-\frac{m}{\hbar\omega^3}(\omega^2 x+\dot{v})^2\right]. \qquad (B.4)$$

Expression (B.4) corresponds to Wigner function (3.5).

Let us check the fulfillment of equation (3.4) for function (B.4). Substitution of function (B.4) into equation (3.4) leads to the equation for function $\gamma(x,v,\dot{v},\ddot{v})$

$$v\frac{\partial\gamma}{\partial x}+\dot{v}\frac{\partial\gamma}{\partial v}+(\ddot{v}-3\omega^2 v)\frac{\partial\gamma}{\partial \dot{v}}-\omega^4 x\frac{\partial\gamma}{\partial \ddot{v}}=0, \qquad (B.5)$$

where

$$\gamma = (v^2+\omega^2 x^2)+\frac{1}{\omega^4}(\omega^2 v-\ddot{v})^2+\frac{1}{\omega^2}(\omega^2 x+\dot{v})^2. \qquad (B.6)$$

The derivatives of function (B.6) are of the form:

$$\gamma_x = 4\omega^2 x + 2\dot{v}, \qquad \gamma_v = 4v - \frac{2}{\omega^2}\ddot{v}, \qquad \gamma_{\dot{v}} = 2x + \frac{2}{\omega^2}\dot{v}, \qquad \gamma_{\ddot{v}} = -\frac{2}{\omega^2}v + \frac{2}{\omega^4}\ddot{v}. \qquad (B.7)$$

Substituting (B.7) into equation (B.5), we obtain the correct identity

$$3\omega^2 xv + 3v\dot{v} + \ddot{v}x - 3\omega^2 vx - 3v\dot{v} - x\ddot{v} = 0. \qquad (B.8)$$

Let us find functions $W^{1,2,3}$ and $W^{1,2,4}$, using relations (2.2)-(2.4), we obtain

$$W^{1,2,3} = m\int_{-\infty}^{+\infty} W^{1,2,3,4} d\ddot{v} = \frac{4m}{(2\pi\hbar_2)^2} e^{-\frac{m}{\hbar\omega}\left[(v^2+\omega^2 x^2)+\frac{1}{\omega^2}(\omega^2 x+\dot{v})^2\right]}\int_{-\infty}^{+\infty} e^{-\frac{m}{\hbar\omega^5}(\ddot{v}-\omega^2 v)^2} d\ddot{v} =$$

$$= \frac{m}{(\pi\hbar_2)^2}\sqrt{\frac{\hbar\omega^5}{m}} e^{-\frac{m}{\hbar\omega}\left[(v^2+\omega^2 x^2)+\frac{1}{\omega^2}(\omega^2 x+\dot{v})^2\right]}\int_{-\infty}^{+\infty} e^{-z^2} dz = m\sqrt{\frac{\pi\hbar_2\omega^3}{\pi^4\hbar_2^4 m}} e^{-\frac{m}{\hbar\omega}\left[(v^2+\omega^2 x^2)+\frac{1}{\omega^2}(\omega^2 x+\dot{v})^2\right]},$$

$$W^{1,2,3} = \sqrt{\frac{m}{\pi^3\hbar^3\omega^3}} e^{-\frac{m}{\hbar\omega}\left[(v^2+\omega^2 x^2)+\frac{1}{\omega^2}(\omega^2 x+\dot{v})^2\right]}, \qquad (B.9)$$

where $z = \sqrt{\frac{m}{\hbar\omega^5}}(\ddot{v}-\omega^2 v)$. Similarly,

$$W^{1,2,4} = m \int_{-\infty}^{+\infty} W^{1,2,3,4} \, d\dot{v} = \frac{m}{(\pi \hbar_2)^2} e^{-\frac{m}{\hbar \omega}\left[(v^2+\omega^2 x^2) + \frac{1}{\omega^4}(\ddot{v}-\omega^2 v)^2\right]} \int_{-\infty}^{+\infty} e^{-\frac{m}{\hbar \omega^3}(\omega^2 x + \dot{v})^2} d\dot{v} =$$

$$= \frac{m}{(\pi \hbar_2)^2} \sqrt{\frac{\hbar \omega^3}{m}} e^{-\frac{m}{\hbar \omega}\left[(v^2+\omega^2 x^2) + \frac{1}{\omega^4}(\ddot{v}-\omega^2 v)^2\right]} \int_{-\infty}^{+\infty} e^{-z^2} dz = \sqrt{\frac{m\omega}{\pi^3 \hbar_2^3}} e^{-\frac{m}{\hbar \omega}\left[(v^2+\omega^2 x^2) + \frac{1}{\omega^4}(\ddot{v}-\omega^2 v)^2\right]},$$

$$W^{1,2,4} = \sqrt{\frac{m\omega}{\pi^3 \hbar_2^3}} \exp\left\{-\frac{m}{\hbar \omega}\left[v^2 + \omega^2 x^2 + \left(\frac{\ddot{v}-\omega^2 v}{\omega^2}\right)^2\right]\right\}, \tag{B.10}$$

where $z = \sqrt{\frac{m}{\hbar \omega^4}} (\omega^2 x + \dot{v})$.

Integrating $W^{1,2,3}$ over the acceleration space results in function $W^{1,2}$:

$$W^{1,2} = m \int_{-\infty}^{+\infty} W^{1,2,3} \, d\dot{v} = m \sqrt{\frac{m}{\pi^3 \hbar^3 \omega^3}} e^{-\frac{m}{\hbar \omega}(v^2+\omega^2 x^2)} \int_{-\infty}^{+\infty} e^{-\frac{m}{\hbar \omega^3}(\omega^2 x + \dot{v})^2} d\dot{v} =$$

$$= m \sqrt{\frac{m}{\pi^3 \hbar^3 \omega^3}} \sqrt{\frac{\hbar \omega^3}{m}} e^{-\frac{m}{\hbar \omega}(v^2+\omega^2 x^2)} \int_{-\infty}^{+\infty} e^{-z^2} dz = m \sqrt{\frac{1}{\pi^2 \hbar^2}} e^{-\frac{m}{\hbar \omega}(v^2+\omega^2 x^2)},$$

$$W^{1,2} = \frac{m}{\pi \hbar} e^{-\frac{m}{\hbar \omega}(v^2+\omega^2 x^2)}, \tag{B.11}$$

where $z = \sqrt{\frac{m}{\hbar \omega^3}} (\omega^2 x + \dot{v})$. A similar result (B.11) is obtained by integrating function (B.10) over space $\ddot{v}$:

$$W^{1,2} = m \int_{-\infty}^{+\infty} W^{1,2,4} \, d\ddot{v} = m \sqrt{\frac{m\omega}{\pi^3 \hbar_2^3}} e^{-\frac{m}{\hbar \omega}(v^2+\omega^2 x^2)} \int_{-\infty}^{+\infty} e^{-\frac{m}{\hbar \omega^5}(\ddot{v}-\omega^2 v)^2} d\ddot{v} =$$

$$= m \sqrt{\frac{m\omega}{\pi^3 \hbar_2^3}} \sqrt{\frac{\pi \hbar \omega^5}{m}} e^{-\frac{m}{\hbar \omega}(v^2+\omega^2 x^2)} = \frac{m}{\pi \hbar} e^{-\frac{m}{\hbar \omega}(v^2+\omega^2 x^2)}. \tag{B.12}$$